\documentclass[useAMS,usenatbib]{mn2e}

\usepackage{natbib}
\usepackage{graphicx}
\usepackage{amssymb}
\usepackage{color}
\usepackage[dvipsnames]{xcolor}
\usepackage{caption}
\usepackage{lipsum,graphicx,multicol}
\usepackage{float}
\usepackage[fleqn]{amsmath}
\usepackage{subcaption}

\title[BAL dust-driven outflows]{ Are BAL outflows powered by radiation pressure on dust? } 

\author[ ]
{W. Ishibashi$^{1,2,3}$\thanks{E-mail: wako.ishibashi@physik.uzh.ch}, A. C. Fabian$^{4}$ and P. C. Hewett$^{4}$
\vspace{0.25cm}
\footnotemark[0]\\
$^{1}$Physik-Institut, Universit$\ddot{a}$t Z$\ddot{u}$rich, Winterthurerstrasse 190, 8057 Z$\ddot{u}$rich, Switzerland \\
$^{2}$Istituto Ricerche Solari (IRSOL), Universit$\grave{a}$ della Svizzera italiana (USI), 6605 Locarno Monti, Switzerland \\
$^{3}$Euler Institute, Universit$\grave{a}$ della Svizzera italiana (USI), 6900 Lugano, Switzerland \\ 
$^{2}$Institute of Astronomy, Madingley Road, Cambridge CB3 0HA 
}

\begin{document}

\pdfminorversion=4

\date{Accepted ? Received ?; in original form ? }

\pagerange{\pageref{firstpage}--\pageref{lastpage}} \pubyear{2012}

\maketitle

\label{firstpage}

\begin{abstract} 
Broad absorption line (BAL) outflows are commonly detected in active galactic nuclei (AGN), but their driving mechanism remains poorly constrained. Here we investigate whether radiation pressure on dust can adequately explain the BAL phenomenon observed in quasars. In the framework of our AGN radiative dusty feedback scenario, we show that dust-driven outflows can reach BAL wind-like velocities ($v \sim 10^4$ km/s) on galactic scales ($r \lesssim 1$ kpc). This is consistent with recent observations indicating that BAL acceleration typically occurs on scales of $\sim 10$ pc, and that the majority of BAL outflows are located at galactocentric radii greater than $\sim 100$ pc. We derive the outflow radial velocity profile and compute the associated outflow momentum rate and kinetic power, which are found to be in agreement with the outflow energetics measured in BAL quasars. Therefore radiation pressure on dust may account for the observed BAL outflow dynamics and energetics. Furthermore, we consider BAL clouds/clumps (leading to a clumpy BAL flow characterised by a wide range of outflowing velocities), and we analyse how the resulting covering factors affect the shape of the absorption line profiles. We conclude that dust-driven BAL outflows may provide a significant contribution to AGN feedback on galactic scales. 
\end{abstract}  



\section{Introduction}
\label{Sect_introduction}

A direct evidence of active galactic nucleus (AGN) feedback in action can be observed in the form of ionised gas outflows, detected as blueshifted absorption lines in the rest-frame ultraviolet spectra of quasars. Broad absorption line (BAL) outflows are characterised by very high velocities ($v \sim 10^3 - 10^4$ km/s) and wide absorption troughs  that are blueshifted with respect to the systemic redshift \citep[e.g.][and references therein]{Weymann_et_1991, Laha_et_2021}. The absorption troughs arise from ionised material: the majority from high-ionisation species, such as C IV and Si IV (HiBAL), and a minority also from low ionisation species, such as Mg II and Al III (LoBAL). A particular sub-class show additional absorption from iron Fe II (FeLoBAL). 

BAL outflows are typically observed in $\sim (10-20) \%$ of optically selected quasars, while the intrinsic BAL fraction can be significantly higher \citep{Hewett_Foltz_2003, Gibson_et_2009, Allen_et_2011}. BAL quasars are usually associated with high luminosities ($L > 10^{46} - 10^{47}$ erg/s), with an enhanced BAL fraction reported in e.g. the WISSH sample of hyper-luminous quasars \citep{Bruni_et_2019}. Recent observational works indicate that the BAL fraction increases with redshift, from $\sim 20\%$ at $z \sim 2-4$ to nearly $\sim 50\%$ at $z \sim 6$ \citep{Bischetti_et_2022, Bischetti_et_2023}. Therefore BAL outflows seem to be common in high-luminosity AGNs, with BAL quasars comprising a significant fraction of the quasar population at high redshifts. 

It has long been recognised that BAL quasars are redder in the rest-frame UV spectra compared to non-BAL counterparts, with LoBAL being more strongly reddened than HiBAL \citep{Weymann_et_1991, Reichard_et_2003, Gibson_et_2009, Zhang_et_2014, Krawczyk_et_2015, Chen_et_2022}. The redder colour observed in BAL quasars is likely associated with the presence of dust grains mixed with the gas, which cause additional dust reddening. This already suggests that the dust may be somehow involved in the BAL phenomenon.  

BAL outflows have been extensively investigated over the past three decades. However, the physical mechanism powering BAL winds is still much debated. What actually drives BAL outflows? A widely adopted model is that of accretion disc winds driven by radiation pressure on UV resonance lines \citep{Murray_et_1995, Proga_et_2000}. In the line-driven scenario, the wind is launched from the inner accretion disc at small radii ($r \sim 0.01$ pc) and is rapidly accelerated to high speeds on sub-pc scales. Another possibility is radiation pressure on dust, which acts on larger scales, beyond the sublimation radius for dust grains ($ r \gtrsim 1$ pc). This suggests the existence of a maximal terminal velocity set by the smallest launching radius, i.e. the dust sublimation radius \citep{Scoville_Norman_1995}. More recently, \citet{He_et_2022} report that the kinematics observed in BAL outflows may be accounted for by dust-driving, with the launch region located close to the dusty torus. 

A key parameter that can help distinguish between the two models is the BAL outflow distance from the centre, i.e. the galactocentric radius $r$. Recent observational measurements indicate that BAL outflows are located at radii of tens of parsecs to a few kiloparsecs, typically at $r \sim 100 \, \mathrm{pc} - 1 \, \mathrm{kpc}$ \citep{Arav_et_2018, Arav_et_2020, Xu_et_2019, Miller_et_2020}. Furthermore, a robust constraint on the BAL location is crucial for the determination of the outflow energetics, such as the momentum flux and the kinetic power. By comparing the observed outflow energetics with the central luminosity output, one can assess the potential contribution of BAL outflows to AGN feedback. 

Here we consider radiation pressure on dust as the driving mechanism for BAL outflows. We examine whether the observed BAL properties can be explained in terms of radiation pressure-driven dusty outflows. We also provide a first quantification of the BAL outflow dynamics and energetics in the framework of our AGN radiative dusty feedback scenario \citep[][]{Ishibashi_Fabian_2015, Ishibashi_Fabian_2018, Ishibashi_et_2021}. 
The paper is structured as follows. We first recall the basics of AGN radiation-driven dusty outflows (Sect. \ref{Sect_RDF_model}), which we now apply to the case of BAL outflows. We analyse the resulting BAL outflow dynamics, and we derive analytic limits for the wind radial velocity profile in the case of fixed-mass shells (Sect. \ref{Sect_BAL_dynamics}). We compute the corresponding BAL outflow energetics in Sect. \ref{Sect_BAL_energetics}  and compare with available observational measurements. The case of expanding shells sweeping up matter from the surrounding environment is treated in Sect. \ref{Sect_exp_shells}. In addition, we consider the case of clumpy BAL outflows, which lead to a range of covering factors and absorption line profiles (Sect. \ref{Sect_clumpy_BAL}). The prospect of such dust-driven BAL outflows is further discussed in the broader context of AGN-galaxy co-evolution over cosmic time (Sect. \ref{Sect_discussion}).  


\section{Outflows driven by radiation pressure on dust}
\label{Sect_RDF_model}

The combination of high luminosity and dust reddening observed in BAL quasars is suggestive of radiatively driven dusty winds. We have previously considered how galactic outflows can be powered by AGN radiation pressure on dust (see e.g.  \citet{Ishibashi_Fabian_2015, Ishibashi_Fabian_2018} for a more detailed discussion of the AGN radiative dusty feedback scenario). 
The equation of motion of the outflowing shell is given by
\begin{equation}
\frac{d}{dt} [M_\mathrm{sh}(r) v] = \frac{L}{c} (1 + \tau_\mathrm{IR} - e^{-\tau_\mathrm{UV}} ) - \frac{G M(r) M_\mathrm{sh}(r)}{r^2} ,  
\label{Eq_motion_shell}
\end{equation} 
where $L$ is the central luminosity and $M_\mathrm{sh}(r)$ is the outflowing shell mass. We assume that the total mass follows an isothermal distribution $M(r) = \frac{2 \sigma^2 r}{G}$ (where $\sigma$ is the velocity dispersion) and we initially consider a fixed-mass shell ($M_\mathrm{sh}(r) = M_\mathrm{sh}$). The infrared (IR) and ultraviolet (UV) optical depths are given by $\tau_\mathrm{IR,UV} = (\kappa_\mathrm{IR,UV} M_\mathrm{sh})/(4 \pi r^2)$, where $\kappa_\mathrm{IR}$=$5 \, \mathrm{cm^2 g^{-1} f_{dg, MW}}$ and $\kappa_\mathrm{UV}$=$10^3 \, \mathrm{cm^2 g^{-1} f_{dg, MW}}$ are the IR and UV opacities, with the dust-to-gas ratio ($f_\mathrm{dg}$) normalised to the Milky Way value. 
The following values are assumed as fiducial parameters of the model: central luminosity $L = 5 \times 10^{46}$ erg/s, shell mass $M_\mathrm{sh} = 10^6 M_{\odot}$, velocity dispersion $\sigma$ = 200 km/s, dust-to-gas ratio $f_\mathrm{dg} = 1 \times f_\mathrm{dg,MW}$, and initial radius $r_0$ = 3 pc. 

The temporal evolution of the outflowing shell is determined by the competition between the outward radiative force and the inward gravitational force (corresponding to the first and second terms on the right hand side of equation \ref{Eq_motion_shell}). 
By equating these two opposite forces, we obtain a critical luminosity that can be considered as a generalised form of the Eddington luminosity 
\begin{equation}
L_\mathrm{E}^{'}(r) =  \frac{2 c \sigma^2 M_\mathrm{sh}}{r} (1 + \tau_\mathrm{IR} - e^{-\tau_\mathrm{UV}} )^{-1} . 
\end{equation} 
The corresponding effective Eddington ratio is given by $\Gamma^{'} = L/L_\mathrm{E}^{'}$. 
To launch an outflow, the central luminosity must exceed this critical luminosity ($L > L_\mathrm{E}^{'}$); or equivalently, the effective Eddington ratio must be greater than unity $\Gamma^{'} > 1$. 

Three distinct physical regimes can be identified depending on the optical depth of the ambient medium: optically thick to both IR and UV, optically thick to UV but optically thin to IR (single scattering limit), and optically thin to UV. 
The corresponding IR and UV transparency radii are defined by $R_\mathrm{IR,UV} = \sqrt{ \frac{\kappa_\mathrm{IR,UV} M_\mathrm{sh}}{4 \pi} }$. 
The effective Eddington ratios in the three optical regimes are given by \citep[cf.][]{Ishibashi_et_2018a}: 
\begin{equation}
\Gamma_\mathrm{IR} 
= \frac{\kappa_\mathrm{IR} L}{8 \pi c \sigma^2 r} 
\propto f_\mathrm{dg} L 
\label{Eq_Gamma_IR}
\end{equation}
\begin{equation}
\Gamma_\mathrm{SS} =  \frac{L r}{2 c \sigma^2 M_\mathrm{sh}}  
\propto \frac{L}{M_\mathrm{sh}}
\label{Eq_Gamma_SS}
\end{equation}
\begin{equation}
\Gamma_\mathrm{UV} 
= \frac{\kappa_\mathrm{UV} L}{8 \pi c \sigma^2 r} 
\propto f_\mathrm{dg} L 
\label{Eq_Gamma_UV}
\end{equation}
We see that the luminosity appears in all three optical depth regimes, and that the effective Eddington ratio is inversely proportional to the shell mass in the single scattering limit. On the other hand, the effective Eddington ratios are independent of the shell mass in the IR-optically-thick and UV-optically-thin regimes, while both $\Gamma_\mathrm{IR}$ and $\Gamma_\mathrm{UV}$ directly scale with the dust opacity and hence dust-to-gas ratio.  


\section{BAL outflow dynamics }
\label{Sect_BAL_dynamics}

\subsection{Radial velocity profile }

\begin{figure}
\begin{center}
\includegraphics[angle=0,width=0.4\textwidth]{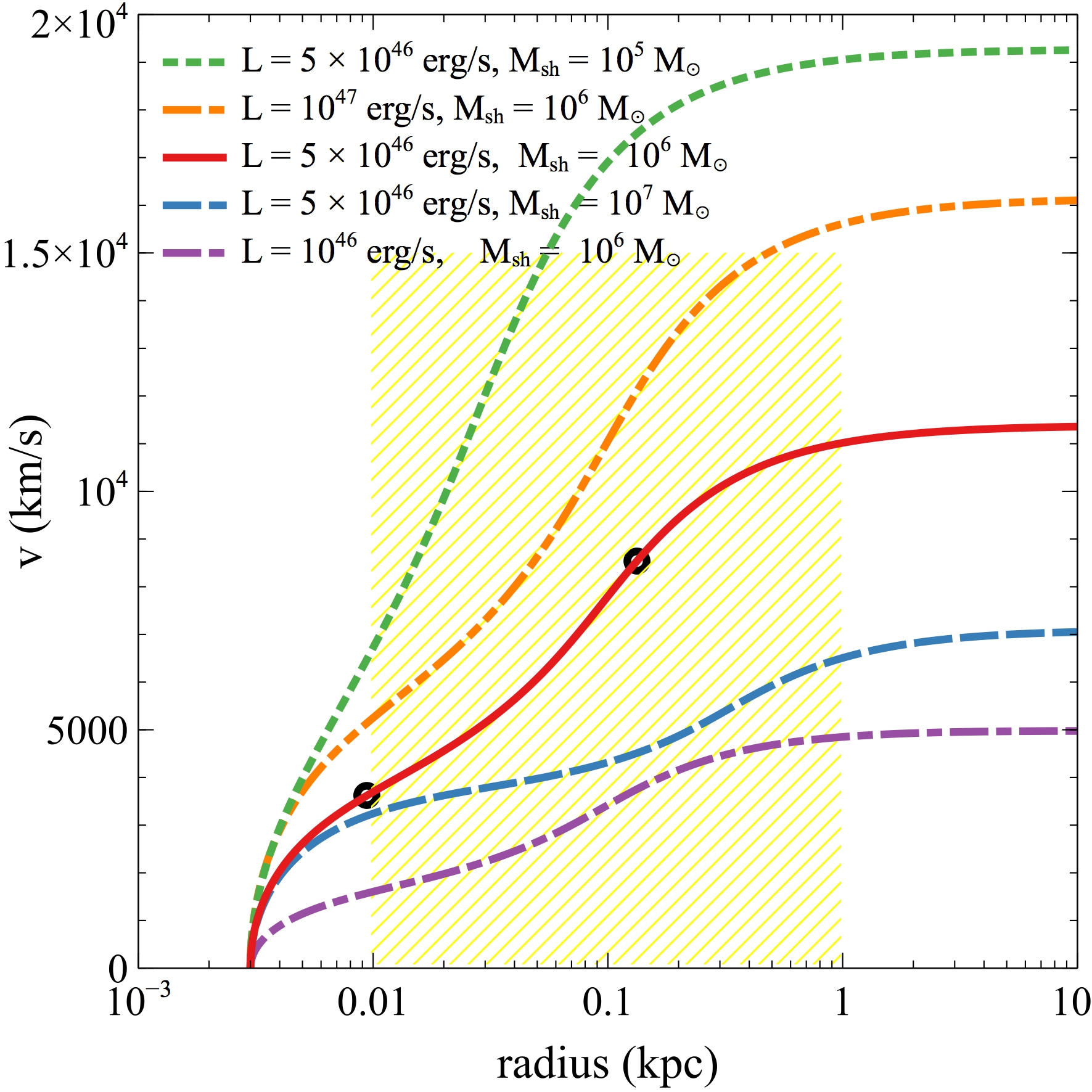} 
\caption{ 
Outflow radial velocity profile $v(r)$ with variations in central luminosity and outflowing shell mass: $L = 5 \times 10^{46}$ erg/s with $M_\mathrm{sh} = 10^5 M_{\odot}$ (green dotted), $M_\mathrm{sh} = 10^6 M_{\odot}$ (red solid), $M_\mathrm{sh} = 10^7 M_{\odot}$ (blue dashed); $L = 10^{47}$ erg/s and $M_\mathrm{sh} = 10^6 M_{\odot}$ (orange dash-dot-dot); $L = 10^{46}$ erg/s and $M_\mathrm{sh} = 10^6 M_{\odot}$ (violet dash-dot). The two black dots mark the location of the IR and UV transparency radii ($R_\mathrm{IR}$ and 
$R_\mathrm{UV}$) for the fiducial case. The yellow shaded area represents the typical outflow velocity range observed in BAL quasars. 
}
\label{Fig_v_r_var_L_Msh}
\end{center}
\end{figure}

Integrating  the equation of motion (equation \ref{Eq_motion_shell}), we obtain the radial velocity profile of the outflowing shell. In Fig. \ref{Fig_v_r_var_L_Msh}, we show the outflow velocity as a function of radius, for different central luminosities and outflowing shell masses. We see that outflows driven by radiation pressure on dust can reach very high velocities, $v \sim$ several 1000 km/s and up to $v \gtrsim 10^4$ km/s, at radial distances of $r \sim (0.01-1)$ kpc. Thus typical BAL wind-like velocities can be attained on galactic scales. This is to be compared to BAL outflow observations with available distance measurements. 

Based on the use of troughs from ionic excited states, \citet{Arav_et_2018} report that at least $\sim 50\%$ of BAL outflows are located at $r > 100$ pc from the central source. This is corroborated by results from a blind survey indicating that $\sim 75\%$ of S IV outflows have $r > 100$ pc \citep{Xu_et_2019}. Indeed, most BAL outflows are observed at large galactocentric distances, with the majority lying between $r \gtrsim 100$ pc and $r \lesssim$ few kpc \citep{Miller_et_2020, Arav_et_2020}. 

From Fig. \ref{Fig_v_r_var_L_Msh}, we observe that the outflow velocity increases with radial distance, reaching terminal asymptotic speeds at large radii ($r \gtrsim 1$ kpc). Thus most of the wind acceleration phase takes place between the launch radius at $\sim$ few pc (beyond the dust sublimation radius) and $\lesssim 100$ pc. This is consistent with the empirical evidence indicating that the BAL acceleration occurs on scales of tens of parsecs \citep{He_et_2022}. Based on BAL trough variability measurements, \citet{He_et_2022} find that the outflow distance increases with velocity --from several pc to a hundred pc-- suggesting that the BAL outflow is typically accelerated around $\sim 10$ pc. Such large radial distances far exceed the typical acceleration region of line-driven accretion disc winds, but are consistent with dust-driven outflows launched from the dust sublimation radius ($R_\textrm{sub} \sim$ few pc for luminous quasars). For given initial conditions, higher outflowing velocities are obtained for higher luminosities and/or lower mass shells (Fig. \ref{Fig_v_r_var_L_Msh}). 

\subsection{ Dependence on physical parameters} 

\begin{figure}
\begin{center}
\includegraphics[angle=0,width=0.4\textwidth]{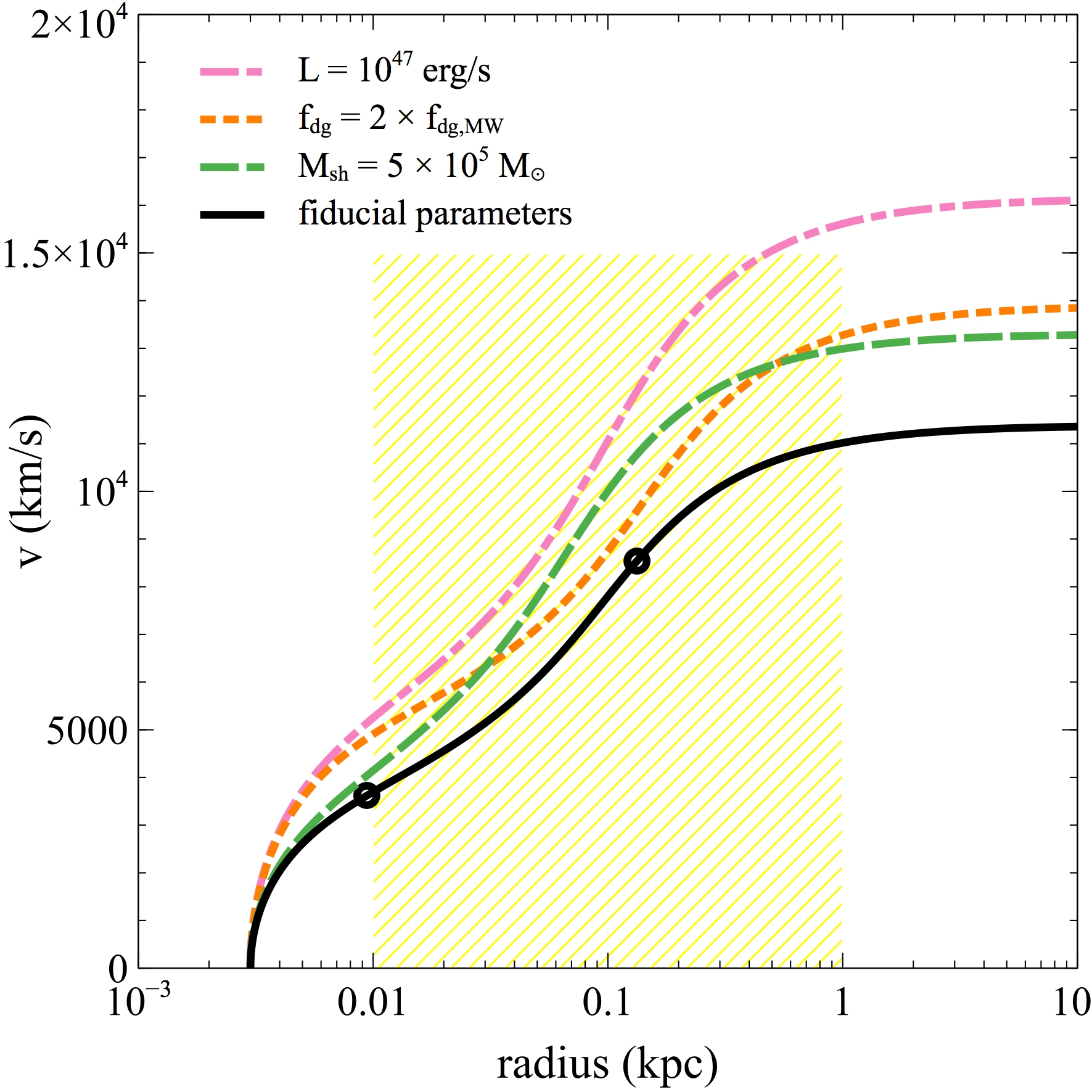} 
\caption{ 
Outflow radial velocity profile $v(r)$ with variations by a factor of two in luminosity, shell mass, and dust-to-gas ratio: fiducial parameters (black solid), $L = 10^{47}$ erg/s (pink dash-dot), $M_\mathrm{sh} = 5 \times 10^5 M_{\odot}$ (green dashed), $f_\mathrm{dg} = 2 \times f_\mathrm{dg,MW}$ (orange dotted). The two black dots mark the location of the IR and UV transparency radii ($R_\mathrm{IR}$ and $R_\mathrm{UV}$) for the fiducial case. The yellow shaded area represents the typical outflow velocity range observed in BAL quasars.
}
\label{Fig_v_r_var_L_Msh_fdg}
\end{center}
\end{figure}

To better understand how the outflow velocity profile depends on the underlying parameters, we plot variations by a factor of two in luminosity, shell mass, and dust-to-gas ratio (Fig. \ref{Fig_v_r_var_L_Msh_fdg}). In general, we see that an increase in luminosity and dust-to-gas ratio, as well as a decrease in shell mass, lead to higher outflowing velocities. 

Breaking down into the three optical depth regimes (introduced in Sect. \ref{Sect_RDF_model}), we note that the central luminosity has the major effect on the outflow velocity, since $L$ appears in all three regimes (equations \ref{Eq_Gamma_IR}-\ref{Eq_Gamma_UV}). Such a luminosity dependence is also consistent with the empirical trend observed in BAL quasars, whereby higher UV luminosity sources reach higher wind velocities \citep{Gibson_et_2009, Bruni_et_2019}. 

The dust-to-gas ratio (appearing in both $\Gamma_\mathrm{IR}$ and $\Gamma_\mathrm{UV}$), enhances the outflow velocity at small radii and large radii (corresponding to the IR-optically-thick and UV-optically-thin regimes, respectively); while it has no effect in the single scattering limit at intermediate radii. In contrast, the shell mass mainly affects the single scattering regime (as $\Gamma_\mathrm{SS}$ scales inversely with $M_\mathrm{sh}$) such that lower mass shells can be accelerated to higher velocities. Hence the shell mass is an important parameter determining the outflow propagation at intermediate radii ($R_\mathrm{IR} \lesssim r \lesssim R_\mathrm{UV}$). Moreover, as the effective Eddington ratio increases with radius in the single scattering limit ($\Gamma_\mathrm{SS} \propto r$), the outflowing shell tends to become increasingly super-Eddington. The highest outflow velocities are obtained for a favourable combination of high luminosity, high dust-to-gas ratio, and low mass shell.

In addition, variations in the initial launch radius can have a significant impact on the velocity profile, with smaller $r_0$ leading to higher outflowing speeds. A lower limit on the initial radius of dust-driven outflows is set by the dust sublimation radius: $R_\textrm{sub} = \sqrt{\frac{L}{4 \pi \sigma_\textrm{SB} T_\textrm{sub}^4}}$, where $\sigma_\mathrm{SB}$ is the Stefan-Boltzmann constant, and $T_\mathrm{sub}$ is the sublimation temperature of dust grains. Assuming $L \sim 10^{47}$ erg/s and $T_\textrm{sub} \sim 1500$ K, the dust sublimation radius is about $R_\textrm{sub} \sim 2$ pc.  

\subsection{ Analytic limits } 

In the case of fixed-mass shells, analytic solutions of the outflow radial velocity profile can be derived from the equation of motion (see also \citet{Ishibashi_et_2021}). Here we provide analytic expressions for the outflow velocity as a function of radius $v(r)$ in two regions, separated by the UV transparency radius $R_\mathrm{UV}$: the UV-optically-thick region at smaller radii (IR+SS regimes) and the UV-optically-thin region at larger radii (UV regime). 

On small scales ($r < R_\mathrm{UV}$), the UV optical depth is much greater than unity ($\tau_\mathrm{UV} \gg 1$), and the UV exponential term can be neglected. The resulting outflow velocity in the UV-optically-thick region is given by 
\begin{equation}
v_\mathrm{I}(r) =  \sqrt{ \frac{2 L}{c M_\mathrm{sh}}(r - r_0) +  \frac{\kappa_\mathrm{IR} L}{2 \pi c} \left( \frac{1}{r_0} - \frac{1}{r} \right) - 4 \sigma^2 \ln \frac{r}{r_0}} , 
\label{Eq_vI}
\end{equation} 
assuming a zero initial velocity ($v_0 = 0$). 

Conversely, at large radii ($r > R_\mathrm{UV}$), the UV optical depth is small ($\tau_\mathrm{UV} \ll 1$), and the exponential term may be approximated as $e^{-\tau_\mathrm{UV}} \sim (1 - \tau_\mathrm{UV}$). The corresponding asymptotic velocity in the UV-optically-thin region is given by 
\begin{equation}
v_\mathrm{II}(r) =  \sqrt{v_\mathrm{I}^2(R_\mathrm{UV}) +  \frac{\kappa_\mathrm{UV} L}{2 \pi c} \left( \frac{1}{R_\mathrm{UV}} - \frac{1}{r} \right) - 4 \sigma^2 \ln \frac{r}{R_\mathrm{UV}}} , 
\label{Eq_vII}
\end{equation} 
where the expression $v_\mathrm{I}(r)$ is evaluated at $r = R_\mathrm{UV}$ (equation \ref{Eq_vI}). 
For fiducial parameters, the outflow velocity at small radii is about $v_\mathrm{I}(r \sim 10 \mathrm{pc}) \sim 3700$ km/s, while the terminal velocity reaches $v_\mathrm{II}(r \sim 1  \mathrm{kpc}) \sim$ 12,000 km/s on kpc-scales. 

\begin{figure}
\begin{center}
\includegraphics[angle=0,width=0.4\textwidth]{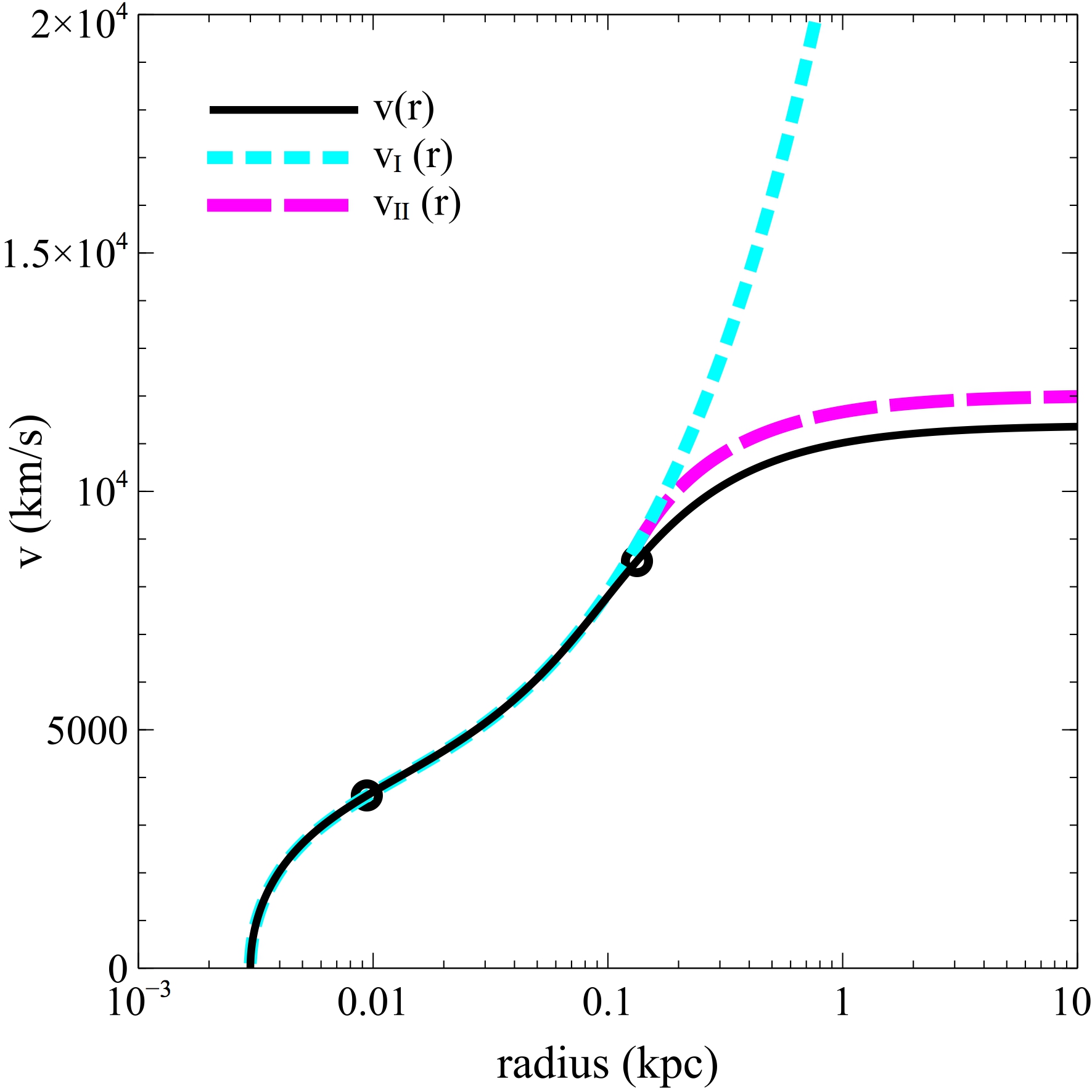} 
\caption{ 
Outflow radial velocity profile: 
comparison between the full numerical solution $v(r)$ (black solid) and the analytic limits, $v_\mathrm{I} (r)$ starting from $r = R_0$ (cyan dotted) and $v_\mathrm{II} (r)$ starting from $r = R_\mathrm{UV}$ (magenta dashed). The two black dots mark the location of the IR and UV transparency radii ($R_\mathrm{IR}$ and $R_\mathrm{UV}$). 
}
\label{Fig_v_r_ana_num}
\end{center}
\end{figure}

Figure \ref{Fig_v_r_ana_num} shows the comparison between the analytic formulae provided in equations \ref{Eq_vI}-\ref{Eq_vII} (coloured curves) and the full numerical solution of the outflow radial velocity profile (black curve). We see that the analytic limits provide an accurate description of the numerical results, in excellent agreement in the respective validity domains. Therefore our analytic solutions may be used for directly estimating the wind velocity (at any given radius) of radiation-driven dusty outflows. 

\section{BAL outflow energetics }
\label{Sect_BAL_energetics}

Knowing the spatial extent of BAL winds, alongside their outflowing velocities, we can now estimate the corresponding energetics. Assuming a thin spherical shell geometry \citep[e.g.][]{He_et_2019, Arav_et_2020}, the mass outflow rate, the momentum flux, and the kinetic power are respectively given by
\begin{equation}
\dot{M} = 4 \pi m_p \mu  \Omega N_\mathrm{sh} r v = \frac{ \mu  \Omega M_\mathrm{sh} v}{r}
\end{equation}
\begin{equation}
\dot{p} = \dot{M} v = 4 \pi m_p \mu  \Omega N_\mathrm{sh} r v^2 = \frac{ \mu  \Omega M_\mathrm{sh} v^2}{r}
\end{equation}
\begin{equation}
\dot{E}_k = \frac{1}{2} \dot{M} v^2  = 2 \pi m_p \mu  \Omega N_\mathrm{sh} r v^3 = \frac{ \mu  \Omega M_\mathrm{sh} v^3}{2r}
\end{equation}
where $N_\mathrm{sh} = M_\mathrm{sh}/(4 \pi m_p r^2)$ is the shell column density, $\mu = 1.4$ is the mean atomic mass per proton, and $\Omega = 0.2$ is the global covering factor \citep{He_et_2019, He_et_2022}. 

The outflow energetics can be further quantified by two related quantities normalised by the AGN radiative output: the momentum ratio and the energy ratio, defined as
\begin{equation}
\zeta = \frac{\dot{p}}{L/c} = \frac{4 \pi m_p \mu  \Omega r N_\mathrm{sh} v^2}{L/c} 
= \frac{\mu \Omega M_\mathrm{sh}v^2}{(L/c) r}
\end{equation}
\begin{equation}
\epsilon_k = \frac{\dot{E}_k}{L} = \frac{2 \pi m_p \mu  \Omega r N_\mathrm{sh} v^3}{L}
= \frac{\mu  \Omega M_\mathrm{sh} v^3}{2 L r} 
\end{equation} 

\begin{figure*}
\begin{multicols}{2}
\includegraphics[width=0.8\linewidth]{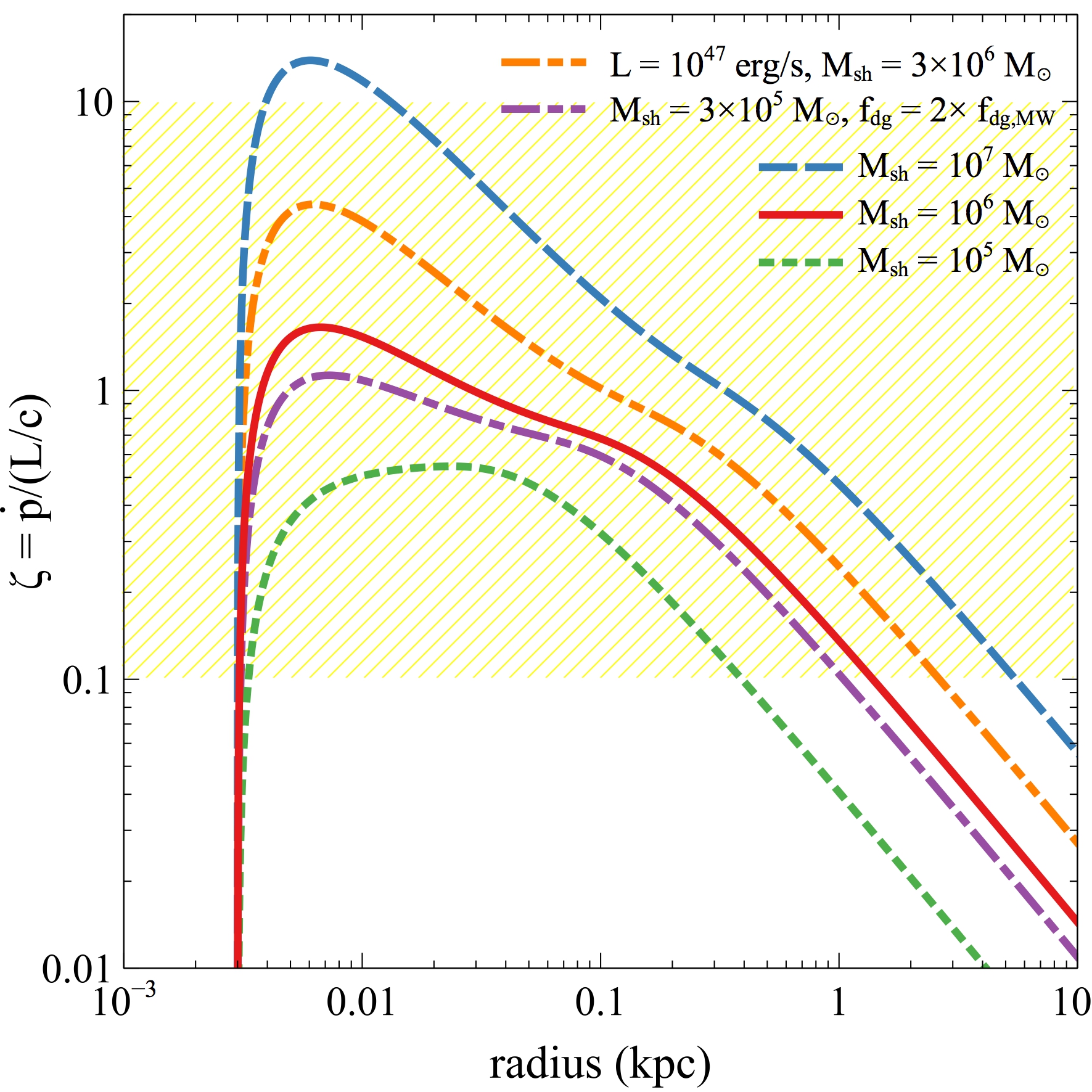}\par
\includegraphics[width=0.8\linewidth]{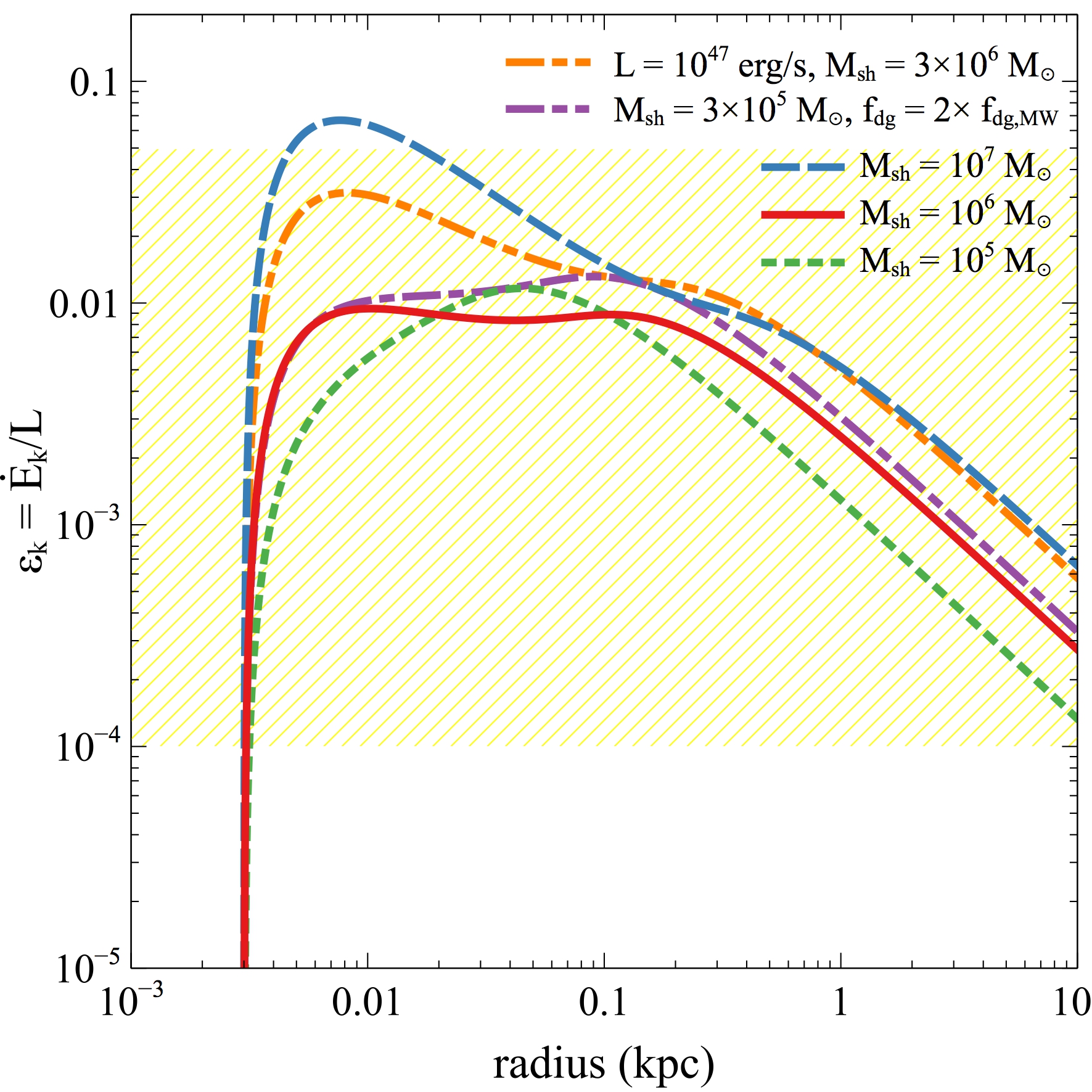}\par 
\end{multicols}
\caption{
Outflow energetics radial profiles: momentum ratio (left-hand panel) and energy ratio (right-band panel) for fiducial values (with $\mu = 1.4$, $\Omega = 0.2$). Variations in shell mass, luminosity, and dust-to-gas ratio: $M_\mathrm{sh} = 10^6 M_{\odot}$ (red solid), $M_\mathrm{sh} = 10^7 M_{\odot}$ (blue dashed), $M_\mathrm{sh} = 10^5 M_{\odot}$ (green dotted); $L = 10^{47}$ erg/s and $M_\mathrm{sh} = 3 \times 10^6 M_{\odot}$ (orange dash-dot-dot); $M_\mathrm{sh} = 3 \times 10^5 M_{\odot}$ and $f_\mathrm{dg} = 2 \times f_\mathrm{dg,MW}$ (violet dash-dot). The yellow shaded area represents the typical range of momentum ratio ($\zeta = \dot{p}/(L/c)$) and energy ratio ($\epsilon_k = \dot{E}_k/L$) observed in BAL quasars. 
}
\label{Fig_energetics}
\end{figure*}

In Fig \ref{Fig_energetics}, we show the radial profiles of the momentum ratio $\zeta$ (left-hand panel) and energy ratio $\epsilon_{k}$ (right-hand panel) of the radiation pressure-driven outflows. We see that the outflow energetics typically span the range $\zeta \sim (0.1 - 10)$ and $\epsilon_{k} \sim (10^{-3} - 0.1)$ on radial scales of $r \sim (0.01 - 1)$ kpc. Observational results indicate that the majority of absorption outflows have momentum ratios in the range $\dot{p}/(L/c) \sim (0.1 - 10)$ and energy ratios of $\dot{E}_k/L \sim (0.001 - 5) \, \%$, with most of the BAL winds being located at galactocentric distances of $r \sim (100-1000)$ pc \citep{Miller_et_2020}. Similarly, the typical kinetic-to-bolometric luminosity ratio is found to be a few percent in a large sample of BAL quasars drawn from the SDSS \citep{He_et_2019}. So the predicted range of momentum ratios and energy ratios in dust-driven outflows can quantitatively account for the energetics observed in BAL quasars. The resulting high energetics ($\epsilon_k \sim \mathrm{few} \, \%$) suggests that BAL outflows are capable of providing an important contribution to AGN feedback on galactic scales. 

\section{ Expanding shells } 
\label{Sect_exp_shells}

Up to now, we have considered the simple case of fixed-mass shells, which allowed us to derive analytic limits for the outflow dynamics (Sect. \ref{Sect_BAL_dynamics}). In more realistic situations, the expanding shell is likely to sweep up mass from the surrounding environment during its outward propagation. The ambient gas density distribution can be parametrised as a power law of radius (cf. \citet{Ishibashi_Fabian_2018})
\begin{equation}
n(r) = n_0 \left( \frac{r}{r_0} \right)^{-\alpha} , 
\end{equation} 
where $n_0$ is the density of the external medium and $\alpha$ is the power law exponent, with $\alpha = 2$ for an isothermal density distribution. 
The expanding shell mass is then given by 
\begin{equation}
M_\mathrm{sh}(r) = M_\mathrm{sh,0} + 4 \pi m_p \int n(r) r^2 dr 
\approx M_\mathrm{sh,0} + 4 \pi m_p n_0 r_0^{\alpha} \frac{r^{3-\alpha}}{3-\alpha}
\label{Eq_exp_shell}
\end{equation} 
where $M_\mathrm{sh,0}$ is the initial mass of the shell. 

\begin{figure}
\begin{center}
\includegraphics[angle=0,width=0.4\textwidth]{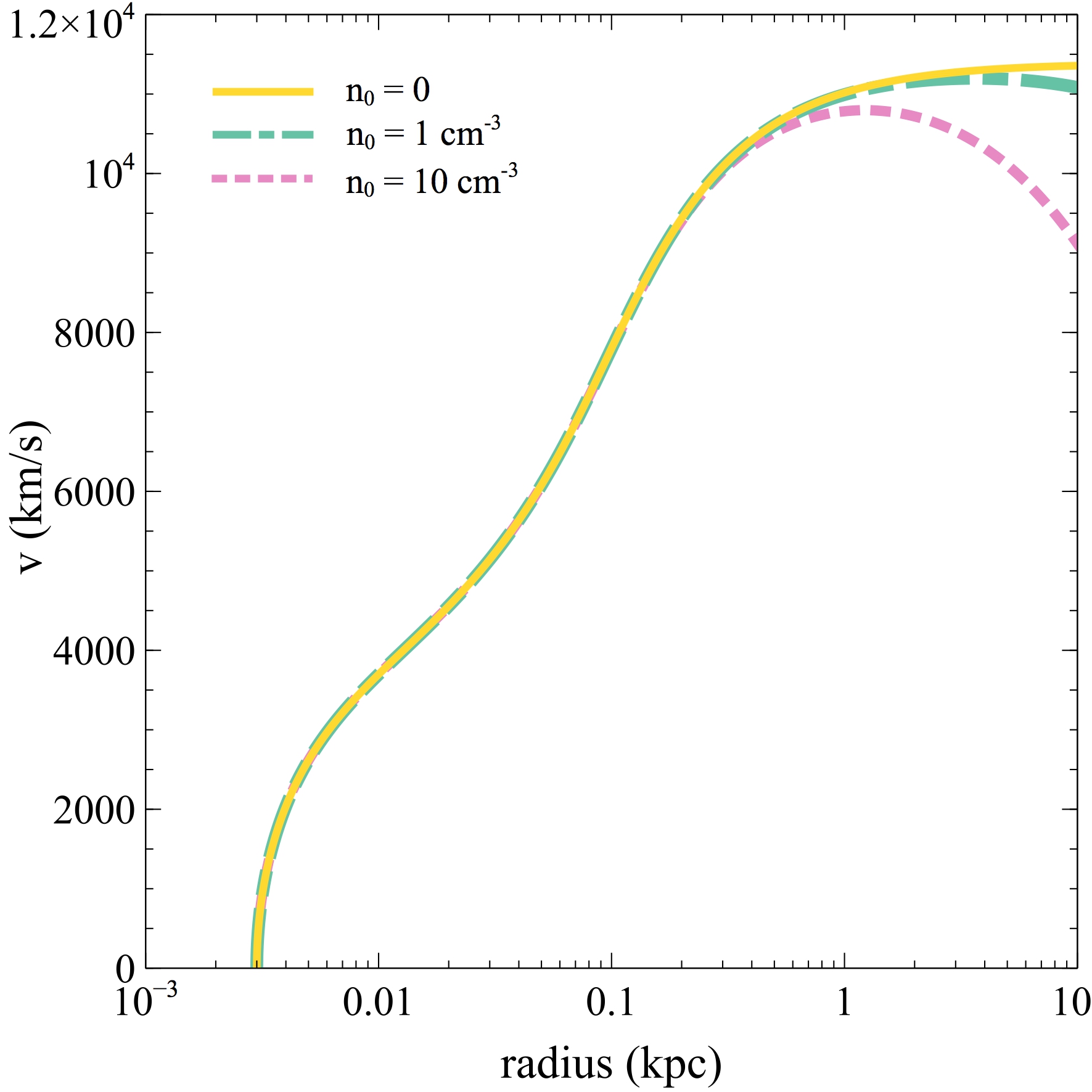} 
\caption{ 
Radial velocity profile of outflowing shells with different external gas densities: 
fixed-mass shell with $M_\mathrm{sh,0} = 10^6 M_{\odot}$ or equivalently $n_0 = 0$ (yellow solid); expanding shells with $n_0 = 1 \, \mathrm{cm^{-3}}$ (cyan dash-dot) and $n_0 = 10 \, \mathrm{cm^{-3}}$ (pink dotted). 
}
\label{Fig_v_r_expsh}
\end{center}
\end{figure}

Figure \ref{Fig_v_r_expsh} shows the radial velocity profile of the outflowing shell for different values of the external gas density $n_0$. As expected, the outflow propagation is slowed down for expanding shells sweeping up mass from the surroundings, and the terminal velocity is lower compared to the case of a fixed-mass shell. As the shell sweeps up more mass, higher gas densities naturally lead to stronger outflow deceleration. We note that the shell velocity is unaffected at small radii ($r \lesssim 1$ kpc), while the growing mass effect becomes dominant at larger radii ($r \gtrsim 1$ kpc). In fact, the outflow dynamics on galactic scales will be determined by the amount of swept-up material, hence the ambient gas density distribution. 

\begin{figure}
\begin{center}
\includegraphics[angle=0,width=0.45\textwidth]{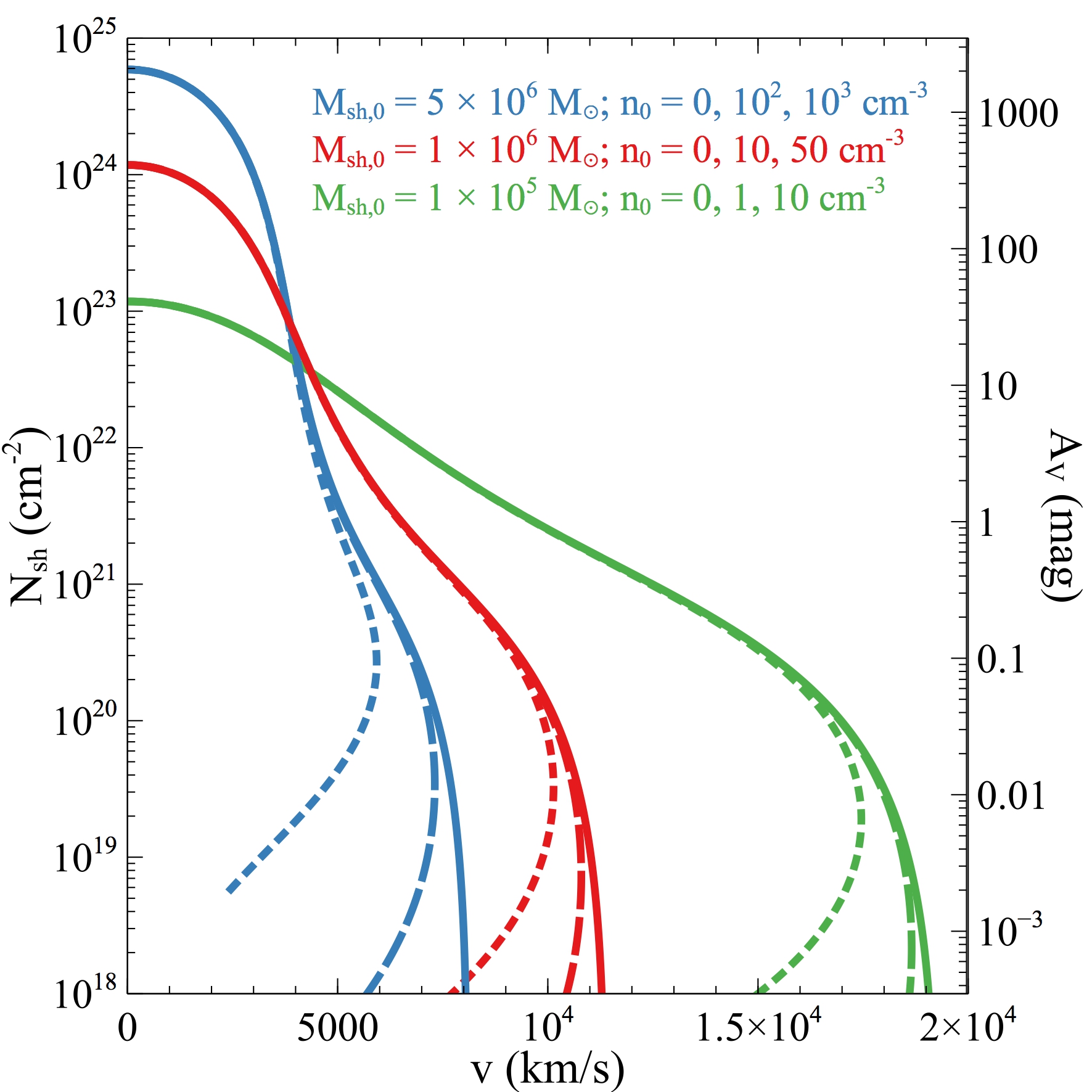} 
\caption{ 
Shell column density $N_\mathrm{sh}$ (left axis) and shell extinction $A_V$ (right axis) as a function of outflowing velocity. Variations in initial shell mass and external gas density: 
$M_\mathrm{sh,0} = 5 \times 10^6 M_{\odot}$, $n_0 = 0, 10^2, 10^3 \, \mathrm{cm^{-3}}$ (blue solid, dashed, dotted); $M_\mathrm{sh,0} = 10^6 M_{\odot}$, $n_0 = 0, 10, 50 \, \mathrm{cm^{-3}}$ (red solid, dashed, dotted); 
$M_\mathrm{sh,0} = 10^5 M_{\odot}$, $n_0 = 0, 1, 10 \, \mathrm{cm^{-3}}$ (green solid, dashed, dotted).
}
\label{Fig_Nsh-AV_v}
\end{center}
\end{figure}

In Fig. \ref{Fig_Nsh-AV_v}, we plot the column density $N_\mathrm{sh}$ of expanding shells (vertical left axis) as a function of the shell velocity $v$. The outflow velocities increase with decreasing column densities, i.e. lower column density shells can be accelerated to higher speeds. Physically, this is simply because lower column density shells have higher Eddington ratios in the single scattering regime ($\Gamma_\mathrm{SS} \propto 1/N_\mathrm{sh}$) and thus attain higher terminal velocities. The column density of the outflowing BAL gas is typically observed in the range $N_\mathrm{H} \sim (10^{21} - 10^{22}) \, \mathrm{cm^{-2}}$, with higher column density gas located at smaller radii \citep{He_et_2022}. This may also be consistent with the empirical anticorrelation between $N_\mathrm{H}$ and BAL distance \citep{Arav_et_2018} and thus the outflow velocity. Figure \ref{Fig_Nsh-AV_v} also shows the plot expressed in terms of the extinction $A_\mathrm{V}$ (vertical right axis) obtained by assuming a linear relation between hydrogen column density and optical extinction, of the form $N_\mathrm{H} (\mathrm{cm^{-2}}) = (2.87 \pm 0.12) \times 10^{21} A_\mathrm{V} (\mathrm{mag})$ \citep{Guver_Ozel_2009, Foight_et_2016}.

\section{Clumpy BAL outflows}
\label{Sect_clumpy_BAL}

\subsection{BAL clouds or clumps} 

Smooth spherical shells subtending $4 \pi$ have been implicitly assumed in the previous sections. We now consider the case of BAL clouds or clumps that subtend only a small fraction of the solid angle from the central source. For instance, the outflowing shell may gradually break up and fragment into clumps, or some individual clouds may be pre-existing. For such clumpy outflows, the wind dynamics will be modified with respect to the spherical shell case, because the temporal evolution of the clump column density becomes decoupled from its radial evolution \citep{Thompson_et_2015}. 
The equation of motion of the clump is given by 
\begin{equation}
\frac{d}{dt} \left[ M_c v_c \right] = \frac{L}{c} \left( \frac{ \pi R_c^2}{4 \pi r^2} \right) \left[ 1 - e^{-\tau_\mathrm{UV}} \right] - \frac{G M(r) M_c}{r^2} , 
\label{Eq_motion_clump}
\end{equation} 
where $M_c$ and $R_c$ are the clump mass and radius, $A_c = \pi R_c^2$ is the cross section of the cloud, $\tau_\mathrm{UV} = \kappa_\mathrm{UV} M_c/A_c$ is the UV optical depth, and the IR term is neglected. %

In the case of clumpy outflows, a new timescale is introduced into the problem: the cloud expansion time $t_\mathrm{exp} = R_c/c_s$, where $c_s$ is the cloud internal sound speed. As the cloud is accelerated outwards by radiation pressure, it will expand laterally in the direction perpendicular to the radial acceleration (due to its finite internal temperature). Assuming that the cloud expands freely, the clump radius will increase as $R_c = R_{c,0} + c_s (t-t_0)$, where $R_{c,0}$ is the clump initial radius at $t_0 = 0$ \citep{Scoville_Norman_1995}. Such lateral expansion implies an increase in the effective cross section of the cloud ($\pi R_c^2$),  leading to an increase in the amount of incident radiation impinging on it. 

As in the case of shells (Sect. \ref{Sect_BAL_dynamics}), the effective Eddington luminosity for the clump is obtained by equating the outward radiative force to the inward gravitational force
\begin{equation}
L_\mathrm{E,c}^{'} 
= \frac{8 \pi m_p c \sigma^2 r N_c}{1 - e^{-\tau_\mathrm{UV}}} , 
\end{equation}
where $N_c = M_c/m_p A_c$ is the column density of the clump.  
The corresponding effective Eddington ratio is $\Gamma_c^{'} = L/L_\mathrm{E,c}^{'}$. 
The condition for the outflow launch ($\Gamma_c^{'} > 1$) requires a critical clump column density of $N_c < \frac{L(1 - e^{-\tau_\mathrm{UV}})}{8 \pi m_p c \sigma^2 r}$. 

We note that the radiative acceleration of the clump in the single scattering regime is inversely proportional to its column density 
\begin{equation}
a_\mathrm{c,SS} 
= \frac{L}{4 \pi m_p c r^2 N_c} ,  
\end{equation}
while in the UV-optically-thin regime, the radiative acceleration becomes independent of the column density and directly scales with the UV opacity
\begin{equation}
a_\mathrm{c,UV} 
= \frac{\kappa_\mathrm{UV} L}{4 \pi c r^2} . 
\end{equation}

\begin{figure*}
\begin{multicols}{2}
\includegraphics[width=0.8\linewidth]{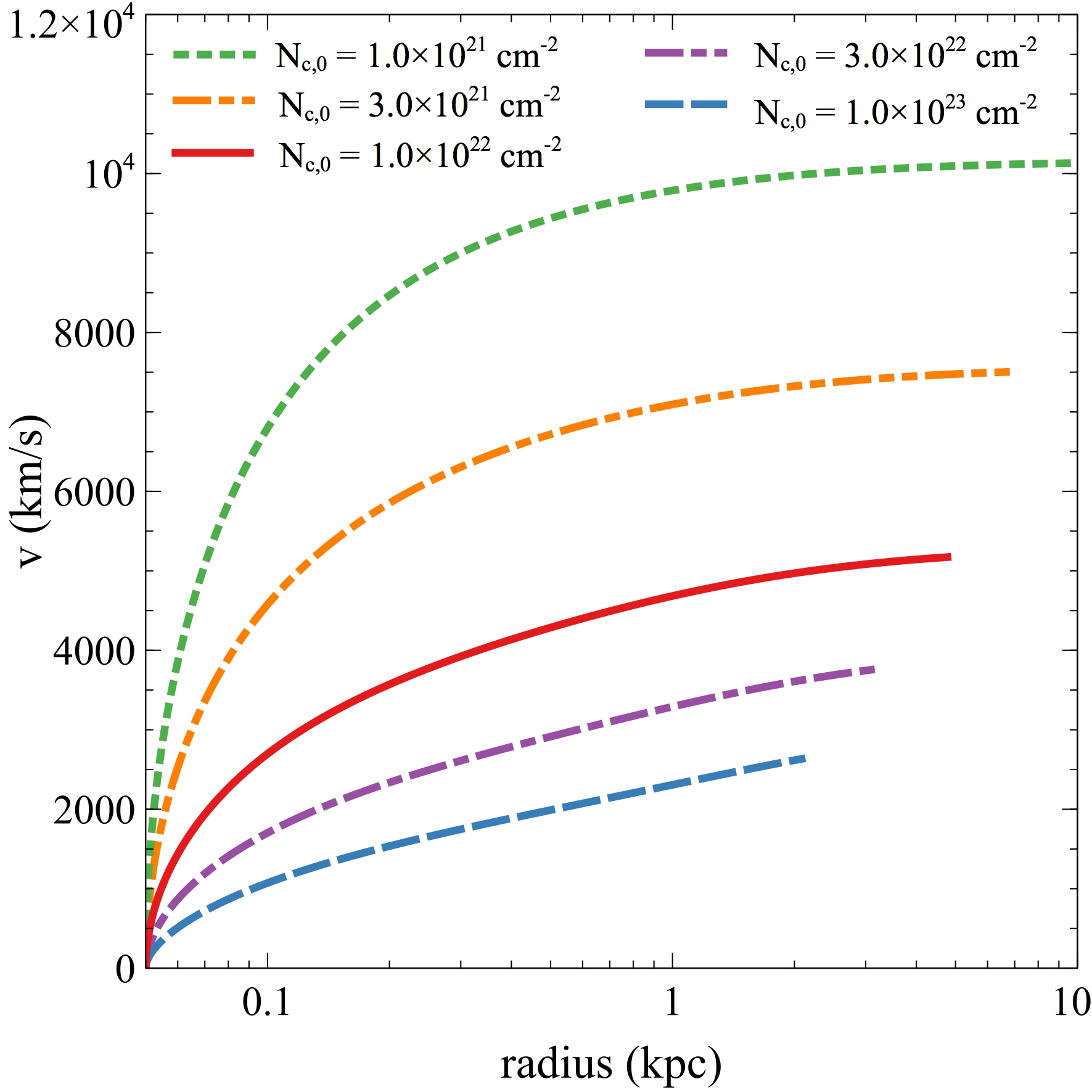}\par
\includegraphics[width=0.85\linewidth]{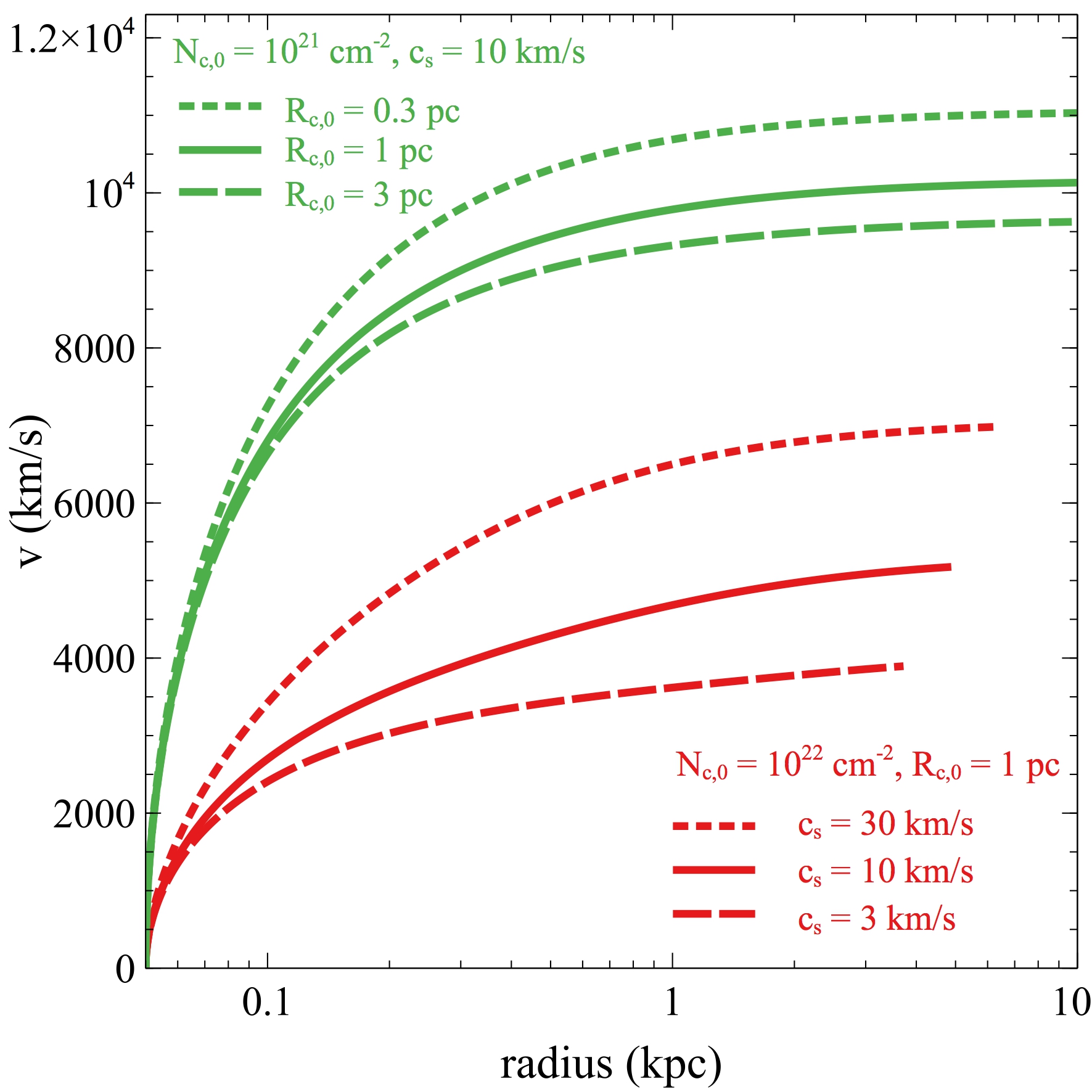}\par 
\end{multicols}
\caption{
Radial velocity profiles of clumpy outflows for fiducial parameters with $R_{c,0}$ = 3 pc, $c_s = 3$ km/s, and $r_0 = 50$ pc. 
Clump radial velocity profiles with variations in initial column density (left-hand panel): $N_{c,0} = 10^{21} \mathrm{cm^{-2}}$ (green dotted), $N_{c,0} = 3 \times 10^{21} \mathrm{cm^{-2}}$  (orange dash-dot-dot), $N_{c,0} = 10^{22} \mathrm{cm^{-2}}$ (red solid), $N_{c,0} = 3 \times 10^{22} \mathrm{cm^{-2}}$ (violet dash-dot), $N_{c,0} = 10^{23} \mathrm{cm^{-2}}$ (blue dashed). 
Clump radial velocity profiles with variations in initial clump radius and internal sound speed (right-hand panel): $N_{c,0} = 10^{21} \mathrm{cm^{-2}}$ with $c_s = 10$ km/s and different initial clump radius: $R_{c,0} = 0.3$ pc (green dotted), $R_{c,0} = 1$ pc (green solid), $R_{c,0} = 3$ pc (green dashed); $N_{c,0} = 10^{22} \mathrm{cm^{-2}}$ with $R_{c,0} = 1$ pc and different sound speed: $c_s = 30$ km/s (red dotted), $c_s = 10$ km/s (red solid), $c_s = 3$ km/s (red dashed).  
}
\label{Fig_v_r_clumpy}
\end{figure*}

Integrating the equation of motion (equation \ref{Eq_motion_clump}), we obtain the radial velocity profile of the outflowing clumps. In Fig. \ref{Fig_v_r_clumpy} (left-hand panel), we show the clump velocity as a function of radius for different values of the initial clump column density $N_\mathrm{c,0}$. We see that lower column density clumps are accelerated to higher velocities and reach larger radii compared to higher column density clumps. This is expected since the radiative acceleration scales as $\propto 1/N_c$ in the single scattering limit, such that lower column density material is more efficiently accelerated. In this picture, lower-$N_c$ clumps might travel ahead of higher-$N_c$ clumps, leading to a range of outflowing velocities. At a given radial location (say $r \sim 1$ kpc), the individual clump velocities can vary from $v \gtrsim 2000$ km/s to $v \lesssim 10^4$ km/s, covering a wide range of outflowing speeds. The velocity spread further increases with increasing radial distance.  

We also analyse how the cloud lateral expansion affects the outflow evolution. Figure \ref{Fig_v_r_clumpy} (right-hand panel) shows the radial velocity profiles varying the initial clump radius $R_{c,0}$ and internal sound speed $c_s$. We observe that an increase in the sound speed leads to higher outflowing velocities (red curves), as higher sound speeds give rise to enhanced cross sections for radiative acceleration. On the other hand, smaller initial clump radii lead to higher outflowing velocities (green curves). The fastest outflows are obtained for a combination of low $N_c$, high $c_s$, and small $R_{c,0}$. Overall, the clump velocities at a given distance from the centre can differ by several thousand km/s, depending on their initial column density, initial size, and internal sound speed. This implies that clumpy BAL outflows are characterised by a wide range of outflowing velocities.  


\subsection{ Covering fraction and absorption line profile }

A BAL system may be formed when outflowing clumps --propagating along the line of sight-- absorb the central radiation. The clump covering factor is given by
\begin{equation}
C_f 
= \frac{\pi R_c^2}{4 \pi r^2} 
= \frac{(R_{c,0} + c_s t)^2}{4 r^2}
= \frac{(R_{c,0} + \frac{c_s}{v} r)^2}{4 r^2} . 
\end{equation}  
If the clump radius remains constant during the outward propagation ($R_c$ = cst), the covering fraction decreases with distance as $C_f \propto 1/r^2$ (geometric dilution). However, if the clump expands laterally as it accelerates outwards, the covering fraction declines more slowly. At large radii, the velocity-dependent covering factor will drop as $C_f \propto 1/v^2$. Therefore larger radii and higher velocities lead to lower covering fractions in accelerating outflows. 

The covering fraction, together with the optical depth, determine the shape of the absorption line profile. 
Following most works in BAL outflows, the normalised flux in the partial covering model is defined as \citep[e.g.][]{Arav_et_2018}
\begin{equation}
F(v) = 1 - C_f(v) + C_f(v) e^{-\tau(v)} , 
\end{equation} 
where $C_f(v)$ is the velocity-dependent covering factor, and $\tau(v) = \kappa_\mathrm{UV} M_c/\pi R_c^2$ is the UV optical depth. 

\begin{figure*}
\begin{multicols}{2}
\includegraphics[width=0.8\linewidth]{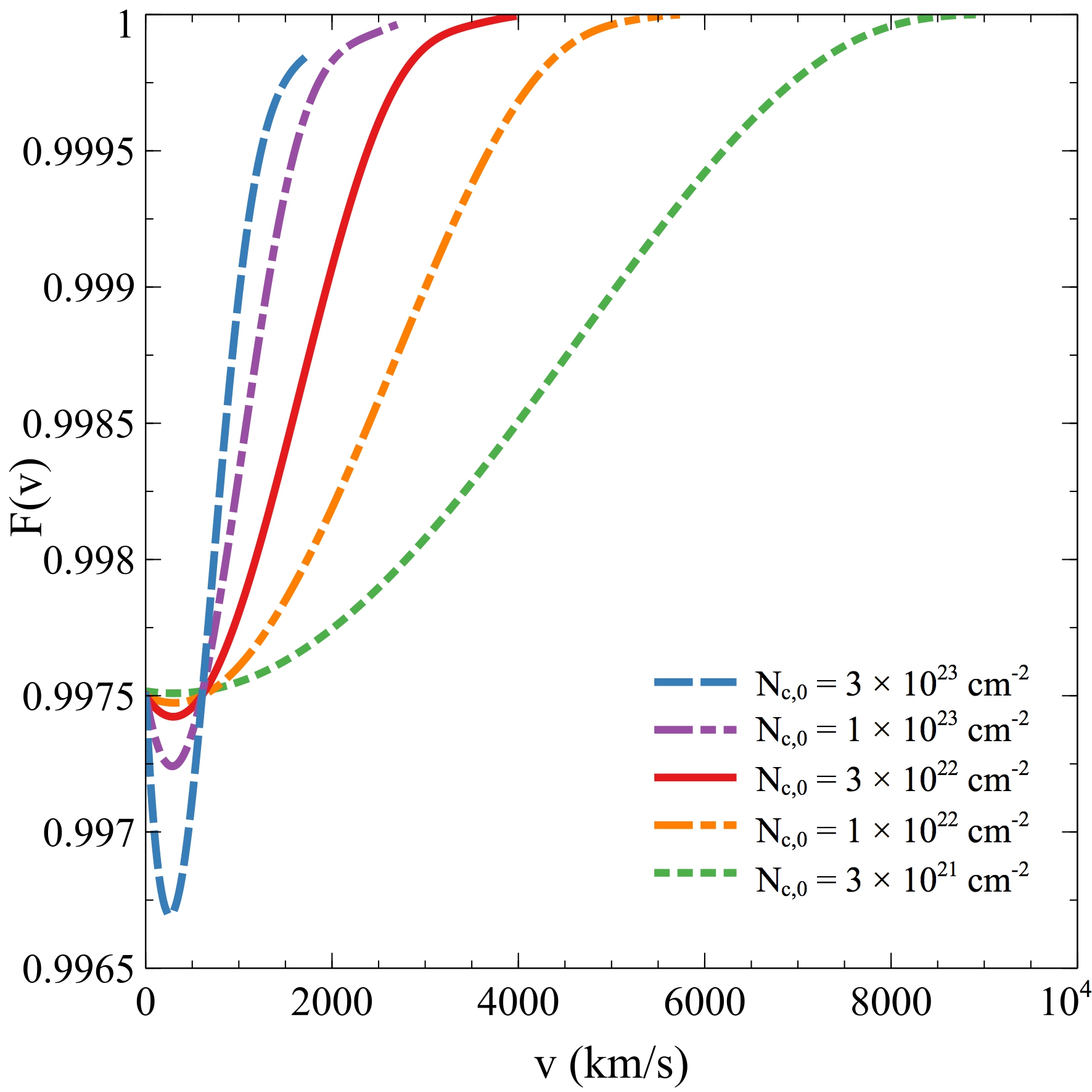}\par
\includegraphics[width=0.8\linewidth]{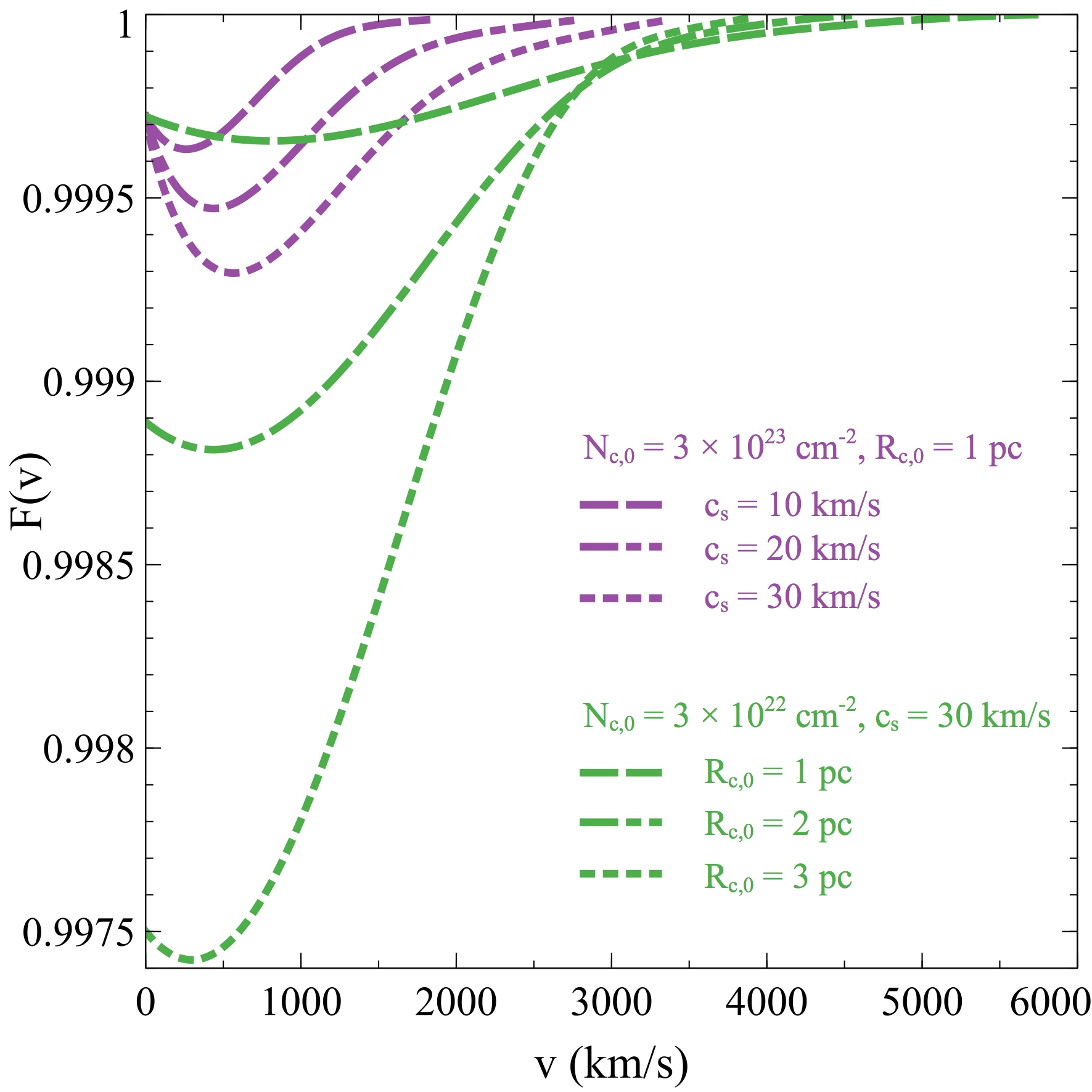}\par 
\end{multicols}
\caption{ 
Normalised flux as a function of outflow velocity for fiducial parameters with $R_{c,0} = 3$ pc, $c_s = 30$ km/s, and $r_0 = 30$ pc. 
Variations initial clump column density (left-hand panel): $N_{c,0} = 3 \times \mathrm{10^{23} cm^{-2}}$ (blue dashed), $N_{c,0} = 10^{23} \mathrm{cm^{-2}}$ (violet dash-dot), $N_{c,0} = 3 \times 10^{22} \mathrm{cm^{-2}}$  (red solid), $N_{c,0} = 10^{22} \mathrm{cm^{-2}}$ (orange dash-dot-dot), $N_{c,0} = 3 \times 10^{21} \mathrm{cm^{-2}}$ (green dotted). 
Variations in initial clump radius and internal sound speed (right-hand panel): $N_{c,0} = 3 \times 10^{23} \mathrm{cm^{-2}}$ and $R_{c,0} = 1$ pc with different sound speeds of $c_s = 10$ km/s (violet dashed), $c_s = 20$ km/s (violet dash-dot), $c_s = 30$ km/s (violet dotted); $N_{c,0} = 3 \times 10^{22} \mathrm{cm^{-2}}$ and $c_s = 30$ km/s with different initial clump radii of $R_{c,0} = 1$ pc (green dashed), $R_{c,0} = 2$ pc (green dash-dot), $R_{c,0} = 3$ pc (green dotted). 
}
\label{Fig_Fv_v_clumpy}
\end{figure*}

Figure \ref{Fig_Fv_v_clumpy} (left-hand panel) represents the normalised flux $F(v)$ as a function of the outflow velocity, for different initial clump column densities. We see that the covering fraction is larger at lower velocities, leading to deep absorption at low speeds, while $C_f$ decreases with increasing velocity. Higher column densities produce deeper absorption troughs with narrower line profiles, whereas lower column densities give rise to shallower troughs with broader line profiles. Meanwhile, larger initial clump radii and higher sound speeds --implying larger covering factors-- lead to deeper absorption line profiles (Fig. \ref{Fig_Fv_v_clumpy}, right-hand panel). The global shape of the absorption line profile is thus determined by the individual clump properties --column density, initial size, and sound speed. The observed BAL trough widths can span thousands of km/s, suggesting that the bulk flow is indeed better described in terms of clumps accelerated to a wide range of velocities rather than single shells \citep{Thompson_et_2015}.  

\section{ Discussion }
\label{Sect_discussion}

While the dynamics and energetics of BAL outflows have been analysed in both individual sources and large quasar samples \citep{Arav_et_2018, Arav_et_2020, Xu_et_2019, Miller_et_2020, He_et_2019, He_et_2022}, the actual BAL acceleration mechanism is still a matter of debate. At least two physical models of BAL outflow driving have been discussed in the literature: accretion disc winds and radiation pressure on dust. Recent observations indicate large galactocentric distances for the BAL location, with the majority of absorption outflows located at radii $r > 100$ pc  \citep{Arav_et_2018, Xu_et_2019, Miller_et_2020}. Furthermore, the BAL acceleration phase likely occurs on scales of tens of pc \citep{He_et_2022}. Such large radii seem to be incompatible with winds originating from the inner accretion disc. Instead, the observational constraints favour radiation pressure on dust --operating beyond the dust sublimation radius-- as the main BAL outflow launching mechanism. 

Here we explicitly show that radiation-driven dusty outflows can be efficiently accelerated to reach BAL-like velocities ($v \sim 10^4$ km/s) on galactic scales ($r \lesssim 1$ kpc). Higher BAL velocities are attained for higher luminosities, lower column densities, and higher dust-to-gas ratios. We also compute the corresponding outflow energetics, and obtain momentum ratios and energy ratios in the range $\zeta \sim (0.1-10)$ and $\epsilon_k \sim (10^{-3}-0.1)$, consistent with available measurements of the outflow energetics in BAL quasars \citep[e.g.][]{He_et_2019, Miller_et_2020}. Therefore BAL outflows, powered by radiation pressure on dust, may provide an important contribution to AGN feedback on galactic scales. 

Observations of BAL outflows indicate typical number densities in the range $n \sim (1.6 \times 10^3 - 2.5 \times 10^5) \, \mathrm{cm^{-3}}$ on radial scales of $\sim 100$ pc to $\sim 4$ kpc \citep{Xu_et_2019}. For a thin spherical shell model, the number density in the outflowing gas can be roughly estimated as $n(r) \sim \dot{M}/(4 \pi m_p r^2 v)$. As the shell moves outwards, it sweeps up mass from the surroundings, and the expanding shell mass grows as $\propto n_0 r_0^{\alpha} \frac{r^{3-\alpha}}{3-\alpha}$ (equation \ref{Eq_exp_shell}), depending on the ambient gas density distribution. For an isothermal full spherical shell ($\alpha = 2$) with fiducial parameters and $n_0 \sim 10^6 \mathrm{cm^{-3}}$, the corresponding number density is about $n \sim (10^3 - 10^5) \, \mathrm{cm^{-3}}$ at $r \sim (0.1-0.01)$ kpc. Larger values of the number density --at larger radii-- may be obtained by considering flatter gas density distributions (e.g. $\alpha = 1,0$). In radiation pressure confinement scenarios, the outflowing gas may be further compressed by radiation pressure \citep{Baskin_et_2014}.
In the more realistic case of a clumpy gas distribution, the cloud number density ($n_c \sim 3 M_c/4 \pi m_p R_c^3$) will depend on the individual clump properties ($M_c, R_c$), leading to a wide range of densities within the outflowing gas.

Relatedly, different ionization states can be expected in BAL outflows, giving rise to high/low ionization lines.  
The ionization parameter is defined as $\xi = L_{i}/(nr^2)$, where $L_{i} = k_i L$ is the ionizing luminosity. Assuming an ionizing fraction of $k_i \sim 0.01-0.5$, the ionization parameter of the fiducial isothermal shell is $\log \xi \sim (0.6 - 2.3) \, \mathrm{erg \, cm \, s^{-1}}$, while the ionization parameter tends to decrease with radius for flatter ambient density distributions ($\alpha < 2$). In reality, in a clumpy inhomogeneous flow, the ionization parameter will depend on the individual cloud properties ($M_c, R_c$), resulting in a broad range of ionization states. This may naturally account for the wide range of ionization levels observed in BAL outflows, with the ionization parameter typically spanning $\log \xi \sim (-0.5 - 2.5) \, \mathrm{erg \, cm \, s^{-1}}$ \citep{Laha_et_2021}. The ionization structure of BAL outflows is further affected by the presence of dust grains. Dust can absorb a significant fraction of the ionizing photons and reduce the ionizing flux, thus facilitating the formation of low ionization species. While high ionization lines may originate in lower density clouds, low ionization lines are more likely produced in high density dusty clumps. 

The distinct spectral features of LoBAL and HiBAL outflows are also reflected in their different dust extinction and reddening properties. LoBAL quasars are found to be significantly redder than HiBAL quasars, with $A_V \sim 0.42 \,  \mathrm{mag}$ and $A_V \sim 0.09 \, \mathrm{mag}$, respectively, based on median composite spectra from SDSS \citep{Chen_et_2022}. The greater reddening is suggestive of a higher dust content in LoBAL quasars; in fact, a particularly high LoBAL fraction is found among dust-reddened quasars \citep{Urrutia_et_2009}. In the framework of the AGN evolutionary sequence \citep[e.g.][]{Sanders_et_1988}, (Lo)BAL quasars may represent a short-lived transition phase from dust-obscured ULIRGs to unobscured luminous quasars. An example is the dusty LoBAL outflow detected in a dust-reddened quasar at $z \sim 2$, which is possibly caught in the act of blowing out its surrounding dust cocoon \citep{Yi_et_2022}. Redder LoBAL quasars may then be associated with an earlier evolutionary phase, subsequently transitioning into bluer HiBAL quasars \citep{Chen_et_2022}.  

Alternatively, dust-driven BAL winds have been interpreted in terms of geometrical configurations, such as in the failed radiatively accelerated dusty outflow (FRADO) model \citep[][and references therein]{Naddaf_et_2023}. In this picture, the BAL phenomenon is only observed when the line of sight lies within the outflowing cone, and is described as an orientation effect. The enhanced reddening of (Fe)LoBAL could then be explained by particular sightlines, e.g. passing close to the edge of the dusty torus \citep{Dunn_et_2015}. 

Direct connections between BAL outflow properties and dust emission have been reported in a number of studies. 
A significant correlation between the outflow strength and the near-IR continuum slope ($\beta_\mathrm{NIR}$, an indicator of the amount of hot dust emission relative to the accretion disc emission) is observed in a large sample of BAL  quasars at $z \sim 2$ \citep{Zhang_et_2014}. More recently, \citet{Temple_et_2021} report that the strength of the $\sim 2 \mathrm{\mu m}$ emission (corresponding to emission from hot dust at the sublimation radius) correlates with the blueshift of the C IV emission line, such that objects with stronger hot dust emission present stronger C IV outflow signatures. Given that larger C IV emission line blueshifts are associated with faster and stronger BAL troughs \citep{Rankine_et_2020}, one may expect causal connections between dust emission and BAL outflow features. A straightforward explanation is that the dust itself is actually providing the opacity for BAL outflow acceleration. 

We note that radiative line-driving also plays an important role in BAL quasars. 
Line-locked CIV absorption systems appear common in Narrow Absorption Line BALs \citep{Bowler_et_2014}. \citet{Lewis_Chelouche_2023} have recently examined the parameter space for line locking to occur, finding that it requires a low column density $N_\mathrm{H} < 10^{19} \mathrm{cm^{-2}}$ and log ionization parameter $U \sim 0$. The relative contribution of the CIV doublet to the total radiation-pressure force for a dusty medium then exceeds one percent (see also \citet{Bowler_et_2014}). This allows Line Locking between clouds along the same line of sight to the nucleus to occur. In the present work, dust driving does the ``heavy lifting'' for clouds to exceed 10,000 km/s, at which point, (see Fig. \ref{Fig_Nsh-AV_v}), $N_\mathrm{H} < 10^{19} \mathrm{cm^{-2}}$, so resonant CIV absorption may be strong enough to lock clouds with a 500 km/s velocity difference together.
Line-locked absorbers can be present in both BAL and non-BAL quasar samples. 

It is interesting to note that BAL and non-BAL quasars are known to share similarities in their emission and continuum properties \citep{Weymann_et_1991}, suggesting that the two sub-classes are drawn from a common parent population \citep{Rankine_et_2020, Temple_et_2024}. In other words, BAL and non-BAL features can be observed in otherwise similar quasars. In our picture, we may envision several AGN outflow stages during the quasar lifetime. A first massive outflow event, sweeping up mass at lower speeds, may clear out the gas from the host galaxy. These slower massive winds without BAL features could be associated with powerful molecular outflows observed in e.g. ULIRGs \citep{Fluetsch_et_2021}. Subsequently, less massive, high-velocity outflows can develop and rapidly propagate into the depleted galaxy, displaying characteristic BAL signatures. The latter may correspond to the fast BAL winds observed in BAL quasars. Overall, radiation pressure on dust provides a promising mechanism for powering BAL outflows across cosmic time. 

BAL outflows are more commonly observed at higher redshifts, with the BAL fraction increasing with redshift \citep[e.g.][]{Allen_et_2011}. The BAL fraction is found to increase up to $\sim 50\%$ at $z \gtrsim 6$, with BAL outflows reaching extreme velocities ($v \gtrsim 0.1 c$) \citep[][]{Bischetti_et_2022}. This suggests an efficient wind acceleration mechanism operating at early times, possibly boosted by radiation pressure on dust. The redshift evolution observed in BAL quasars cannot be accounted for by differences in central luminosity and accretion rate. The higher BAL fraction in the early cosmic epoch may be explained by wider angle outflows in the orientation scenario, or longer blowout phases in the AGN evolutionary sequence \citep{Bischetti_et_2023}. In either case, widespread and powerful dust-driven BAL outflows might play a critical role in shaping black hole-galaxy co-evolution in the early Universe.  


\section*{Acknowledgements }

We thank the anonymous referee for a constructive report. ACF acknowledges early work on the topic by Naoki Arakawa.


\section*{Data availability}

No new data were generated or analysed in support of this research.


\bibliographystyle{mn2e}
\bibliography{biblio.bib}

\label{lastpage}

\end{document}